\documentclass[letterpaper,twocolumn,twoside,slac]{revtex4}
\usepackage{times}
\usepackage{graphicx}
\usepackage{fancyhdr}
\begin{document}

\title{DATA MANAGEMENT FOR PHYSICS ANALYSIS IN PHENIX (BNL, RHIC)}

%

\author{Barbara Jacak}
\affiliation{SUNY at Stony Brook, NY, 11794, USA}

\author{Roy Lacey}
\affiliation{SUNY at Stony Brook, NY, 11794, USA}

\author{Dave Morrison}
\affiliation{BNL, NY, 11973, USA}

\author{Irina Sourikova}
\affiliation{BNL, NY, 11973, USA}

\author{Andrey Shevel}
\affiliation{SUNY at Stony Brook, NY, 11794, USA}

\author{Qiu Zhiping}
\affiliation{SUNY at Stony Brook, NY, 11794, USA}

\begin{abstract}

Every year the PHENIX collaboration deals with increasing volume of data (now about 1/4 PB/year).  
Apparently the more data the more questions how to process all the data in most efficient
way. In recent past many developments in HEP computing were dedicated to the production environment. Now
we need more tools to help to obtain physics results from the analysis of
distributed simulated and experimental data. Developments in Grid architectures gave many
examples how distributed computing facilities can be organized to meet physics analysis needs. We feel
that our main task in this area is to try to use already developed systems or system components in
PHENIX environment.

We are concentrating here on the followed problems: 
file/replica catalog which keep names of our files, 
data moving over
WAN,
job submission in multicluster environment. 

PHENIX is a running experiment and this fact narrowed our ability to test new software on the
collaboration computer facilities.  We are experimenting with system prototypes at State
University of New York at Stony Brook (SUNYSB) where we run midrange computing cluster for physics
analysis \cite{sunysb}. The talk is dedicated to discuss some experience with Grid software and
achieved results.

\end{abstract}

\maketitle

\thispagestyle{fancy}

\section{Introduction}

PHENIX \cite{phenix} is a large, widely spread collaboration where many organizations from many
countries participate (12 nations, 57 institutions, about 400 collaborators, about 1/4 PB is planned
to be produced in the current year). The collaboration is in third year of data taking. Distributed
data (experimental and simulated), distributed computing facilities and distributed users is our
reality.

Before discussing technical decisions we have to emphasize a range of common things and
features in Grid computing for HEP which are important for further discussion.

First of all we paid attention that many systems for physics analysis are already in developing or
prototyping stage \cite{sam, grappa, chimera, alien, nordugrid, magda}. We recognized that using the
Globus tools \cite{globus} is now standard de facto for distributed systems in HEP.  
The last thing was emphasized many times on CHEP2003 \cite{chep2003}.

In more detail we could see the following common components (on all projects):

\begin{itemize}

\item using Globus Security Infrastructure (GSI);

\item using the replica/file catalog for files distributed around the globe, however
different projects use different cataloging engines: Globus Replica Catalog \cite{rc}, other
types of file catalogs;

\item using the GridFTP over WAN in conjunction with other data moving protocols
\cite{bbftp, cpftp};

\item job submission with Globus tools;

\item using the concept of virtual organization.

\end{itemize}

That is not surprising because every collaboration (PHENIX is not exclusion) needs 
several common tools:

\begin{itemize}
\item data moving over WAN;
\item job submission from distributed users to distributed computing clusters;
\item monitoring tools.
\end{itemize}

\noindent
On other hand the character of using Globus middleware is
different in different projects.
If someone tries to see in deeper details - a lot of differences between
collaborations could be discovered. Those differences are rooted in the style of physics 
measurements
which in turn depends on the details of the physics experiment, style of collaboration work,
existing computing resources, prehistory of computing infrastructure and
many other details.  In other words there is no evidence that concrete large 
and sophisticated system
built on top of Grid middleware might be used in a different collaboration 
without reasonable adaptation or redesigning.

That means we have to discuss briefly PHENIX computing environment for physics analysis.

\section{PHENIX computing environment for physics analysis}

PHENIX has several computing facilities for physics analysis:

\begin{itemize}

\item RHIC Computing Facility - RCF \cite{rcf} - main computing facility for PHENIX;

\item CC-J \cite{ccj};

\item midrange cluster at SUNYSB \cite{sunysb};

\item also there is a range of other computing facilities which is used for PHENIX physics analysis at
several member institutes of PHENIX.

\end{itemize}

It is assumed PHENIX will have more computing facilities in future.

By taking a look at PHENIX member list it is clear that data distribution inside collaboration is not
trivial.  Even limited task to know the files are (in which site and location) is not possible without
file/replica catalog (or file catalog).

\section{File cataloging}

In general file catalog is required to keep locations of all data files in the collaboration. There
was a range of various decisions \cite{fatmen, magda} concerning file catalog.

Architecturally we tested two level {\it file cataloging engine}: distributed PHENIX file
catalog \cite{frog, argo} is located at BNL and at remote universities and local SUNYSB file catalog 
based on MAGDA \cite{magda}.

All instances of the catalog are interconnected by special replication process.

Technical description for central PHENIX file catalog was available in a different presentation on
CHEP-2003 \cite{argo-chep}.

At the same time it was recognized that remote universities need some cataloging facilities for 
internal use.

\section{Local cataloging and data moving at SUNYSB}

An adapted version of MAGDA \cite{magda} was implemented and used at SUNYSB as local (at SUNYSB)  
cataloging facility \cite{magdaf}. With time, it became clear that it is suitable and important
to keep the information about many files (physics data and other type of files: scripts,
papers, etc.). Part of this information (information about files with physics data) is
propagated to the central PHENIX file catalog.

There are several tools on top of adapted MAGDA at SUNYSB which are most interesting for end users.
First of all there are web pages \cite{magdaf} with detailed information where and which files are
available.  Another tool is the set of scripts to link locally available data files to the local
directory.

As described before, SUNYSB MAGDA catalog has replicated subset of central PHENIX file catalog.
Periodically cronjob starts the scripts to transfer the information about part of the files from
central PHENIX file catalog to MAGDA catalog and back, part of information from MAGDA
catalog is to be copied to central PHENIX file catalog.

In this way it is possible to keep detailed information about files which are interesting for SUNYSB.

\subsection{To link local data files}

In this context we mean that commands below 
create the soft link for locally available files.
Also information about linked files can be seen at the web pages \cite{magdaf} if the 
panel {\it Used Files} will be clicked.

\begin{itemize}

\item \verb'ma_LinkLocalFile' {\it lfn} - to link locally available file {\it lfn} otherwise special
completion code will be returned;

\item \verb'ma_LinkFileList' {\it list}   - to link locally available files from the {\it list};

\item \verb'ma_LinkFileSubstr' {\it substring} - to link all files which names are containing the
{\it substring};

\item \verb'ma_ShowLinkedFiles'        - to display all files linked by current user;

\item \verb'ma_ReleaseFile' {\it lfn}  - to release the file {\it lfn};
\item \verb'ma_ReleaseHereFiles'    - to release all files in current directory;
\item {\bf \verb'ma_ReleaseAllFiles'}     - to release all files earlier linked by current user.

\end{itemize}

When data file names are released the following steps are performed for every file:

\begin{itemize}

\item the appropriate soft link is deleted;

\item the appropriate record is deleted from the MAGDA database; that means this record will not 
appear anymore in output of the command \verb'ma_ShowLinkedFiles' and in statistics delivered
on the web pages.

\end{itemize}

The information to the catalog MAGDA is coming from special spider scripts which are running on
required {\it sites} (in our case there are 3 sites where spiders are running).  On most sites a
spider is started once a day or even once a week if the information is not 
changed often.

\subsection{Data moving}

File moving over WAN is done at SUNYSB through use of adapted MAGDA and through
an alternative way - by a script. With MAGDA user could go to the web site \cite{magdaf} and
describe required {\it site}, {\it host}, {\it location} (actually the exact directory path) and
{\it collection} (collection of files).  After that it is possible to describe the {\it task}
for data transfer (with web pages on \cite{magdaf}).  Real data moving is possible
after activating the {\it task} by using the web pages \cite{magdaf}. At night cronjob
will perform all the activated {\it tasks}.

For the user convenience the script {\bf gcopy} was developed to copy the files over WAN. The script
uses cluster descriptions which is discussed in chapter {\it Job submission} of this paper. The usage
of the script:

\noindent 
{\bf gcopy} {\it FromSite}:{\it FromLocation} {\it ToSite}:[{\it ToLocation}] 
\verb'\'   

[{\it substring}] 

The parameters {\it FromSite} and {\it ToSite} are names of Globus
gateways. {\it FromLocation} and {\it ToLocation} are exact directory paths. The 
parameter {\it substring} with wild-cards may be  used to select file names in directory {\it 
FromLocation} to be copied.

Technically the file transfer is performed with GridFTP protocol (command {\bf globus-url-copy}). The
feature {\it third party transfer} is used the feature because the host where the file transfers are
started is not the part of any computing cluster. Default number of threads for data transfer is 5.
Due to a range of network routers (6 or so) between SUNYSB and BNL and due to other causes we see a
significant difference in network connectivity speed during a day (a factor of 2). The best throughput
we saw was about 7 MBytes/sec.  This maximum throughput could be reached with different number of
transfer threads and different size of network window at different time.

Taking into account those facts we conclude that it is difficult to predict what time the data
transfer between BNL and SUNYSB will take. This is true for relatively large portion of data (0.2 TB
and more).

Finally, it is much better to be sure that your data are available locally on the computing cluster
where you plan to submit the jobs before job submission.


\section{Job submission}

In a distributed environment, it is often effective to use several computing clusters in
different sites to get enough computing power or to load available clusters more evenly.

We have to emphasize that nobody in the collaboration needs computing power as it is.  
Physicists have a need to use {\bf {\it qualified}} computing power. That means such a computing
cluster where all PHENIX software is already installed and ready to be used. In further
discussion we will assume the following:

\begin{itemize}

\item all required application software has been installed; 

\item required physics data are already locally available or if you plan to do a simulation you have
enough disk space for the output simulated data;

\item all Globus tools have been deployed;

\item users have already the certificates to use Globus Secure Infrastructure (GSI) as well.

\end{itemize}

As already mentioned, the use of Globus infrastructure for job submission is common place now.  
At the same time till last autumn (2002) we had some difficulties with Globus toolkit (GT)  
(especially with data transfer). It was decided to create a light weight testbed with
minimum functionality of GT, with minimum efforts, and with minimum time for implementation which
could be tested in real environment where conditions are close to production reality. To
do that a simple set of scripts was developed.

Our script set (set of wrappers on top of GT 2.2.3) for job submission and job output retrieval
is deployed at client side. About 30 scripts were developed with total number of lines
about 2000. Several of them are most significant for users.

\begin{itemize}

\item {\bf GPARAM} - script to describe the configuration: number of computing clusters, names of 
Globus gateways, other information; in addition the script \$HOME/.gsunyrc (same meaning as {\bf 
GPARAM}) is used to keep local values for
current account;

\item {\bf gproxw} - to create Globus proxy for a week;

\item {\bf gping} - to test availability of all clusters described in {\bf GPARAM};

\item {\bf gping-}{\it M} to test availability of a desired cluster (here {\it M} is
suffix to denote a cluster: {\bf s} - for cluster at SUNYSB \cite{sunysb}, {\bf p} - for PHENIX at BNL 
\cite{rcf}, 
{\bf unm} for
cluster at University of New Mexico \cite{unm}); 

\item {\bf grun-}{\it M} to perform one command (script) on a desired cluster (see remark
to {\bf gping-}{\it M});

\item {\bf gsub-}{\it M} {\it job-script} - to submit the {\it job-script} to a desired computing cluster
(see remark to {\bf gping-}{\it M});

\item {\bf gjobs-}{\it M} [{\it parameter}] - to get the output of command {\bf qstat} on a
desired cluster adn {\it parameter} is parameter to {\bf qstat} (see remark
to {\bf gping-}{\it M});

\item {\bf gsub} {\it job-script} - to submit the {\it job-script} to less loaded computing 
cluster;

\item {\bf gsub-data} {\it job-script} {\it file} - to submit the {\it job-script} 
to the cluster where file {\it file} is located; if the file {\it file} has replica on all 
clusters - to submit to less loaded cluster. First of all the file location will be tested through 
local (on site) file catalog.

\item {\bf gget} {\it job-id} - to get output from accomplished job {\it job-id}, if parameter 
is missing then last submitted job will by taken into account.

\item {\bf gstat} {\it job-id} - to get status of the job {\it job-id}, if parameter is missing 
then last submitted job will be taken into account. 

\end{itemize}

If submitted job generates some files they will be left on the cluster where the job was performed. 
The files could be copied to a different location by data moving facilities described in previous 
section of the paper.

The meaning of {\bf\it less loaded cluster} is important.

\subsection{Less loaded cluster}

We need to know on which cluster the expected execution time for the job is minimum.  
Unfortunately this task in an unstable environment has no simple and precise
solution. The estimate becomes worse if the job runs long time (many hours for instance). All
estimates might be done only on some level of probability.

That was the reason why we took for a prototype a simple algorithm to determine {\bf\it less loaded
cluster}. In principle that choice reflects our hope that situation with job queues will not be
changed fast. Values from queuing system (the answer from the command {\bf qstat}) are used in 
algorithm.
The estimate algorithm uses also a priori information about the average relative power of a node
in the cluster. In our configuration we use two clusters:  average computing power for the
node at SUNY was determined as 1, average computing power at BNL was determined as 2 (the machine
at BNL was twice faster). Another parameter that is used the maximum number of jobs which may be in 
run
stage at SUNY and at BNL. All those parameters are  assigned in the cluster description script
{\bf PARAM}.

Just before starting the job the scripts {\bf gsub, gsub-data} will gather the information about 
real status of queues on every cluster described in the configuration (it is done with Globus command 
{\it globus-job-run} which gets the answer from {\bf qstat}). After that the following value for each
cluster  is 
calculated:

L = [({\small number of jobs in run stage}) + ({\small number of jobs in wait stage}) - 
({\small maximum jobs in run stage})] / ({\small relative computing power})

The cluster with a minimum value of L is considered as {\bf\it less loaded cluster}. Of
course it is only an estimate of the reality. Some peculiarities of a dispatching policies
in local job manager (LSF, PBS, other) could make the above estimate wrong. However in most simple
cases it gives the right direction.

More sophisticated algorithms might be discussed separately.

The current client part is really simple.

\subsection{How to use described scripts on client side}

We assume the client side (usually desktop or laptop) has a stable IP address and can be seen in the
Internet. This feature is mandatory to use all client Globus stuff (we used GT-2.2.3 and GT-2.2.4 on
client side).

\begin{itemize}

\item Deploy the Globus toolkit.

\item
Copy all scripts from \cite{gsuny}. After that please become root and do
\begin{verbatim}
tar zxvf gsuny.tar.gz
cd GSUNY
./setup
\end{verbatim}

All the scripts will be at the directory /usr/local/GSUNY/. To make them available please add this
directory to the \$PATH environment variable.

\end{itemize}

Now the client is ready to use almost all the commands (excluding {\bf gsub-data} which requires 
an additional package \cite{suny-magda}).

First command has to be 
{\bf gproxw}, it creates the Globus proxy for a week.
You could start {\bf gping} after that. The command {\bf gping} will show existing configuration
(number of clusters, Globus gateways, other parameters).

Now you could submit the jobs to described clusters.  If you submit many jobs (several tens or
hundred) it may happen that they will run on different clusters. The job submission scripts use
special logs to remember where the jobs were submitted and which Globus commands in multicluster
environment were performed.

\subsection{User log files in multicluster environment}

In order to keep the trace of user commands in Globus context and job submission
several log files have been created and used:

\begin{itemize}

\item \verb'$USER/.globus/CLUSTERS/commands' - file contains list of performed Globus commands in format 
{\it date/time command parameters} ;

\item  \verb'$USER/.globus/CLUSTERS/jobs' - file contains job ID (with name of globus gateway) in format
{\it date/time jobID}

\item \verb'$USER/.globus/CLUSTERS/DataTransLogs' - directory contains several files to keep trace 
of data moving over WAN.

\end{itemize}

Almost all of mentioned scripts add some records to the above log files.

The logs are very valuable for many reasons: debugging, statistics, etc. They are also important
because the set of clusters which is used by a user may be different.

\subsection{Changing the set of involved clusters}

The technical adding of a new computing cluster consists of
several simple steps for every/only client computer:

\begin{itemize}

\item to change the parameters you can edit the script /usr/local/GSUNY/GPARAM (for system wide
parameters) or to edit the script \$USER/.gsunyrc (for current account);

\item to talk to cluster authority to add your account on new cluster to the file 
/etc/grid-security/grid-mapfile on new cluster Globus gateway.

\item to prepare three scripts and put them into the directory /usr/local/GSUNY/

\begin{itemize}

\item script to ping the new cluster (please see {\bf gping-p} as an example);

\item script to submit the job (please see {\bf gsub-p} as an example);

\item script to get the new cluster load level (please see {\bf gchk-p} as an example).

\end{itemize}

\end{itemize}

After that you can use the cluster under your own account from the desktop where you changed the scripts.

If the cluster has to be from the configuration, you have to edit the script
/usr/local/GSUNY/GPARAM accordingly: it is possible to delete the description of the cluster and
change the number of clusters.

If you have to change the cluster (to use new cluster instead old one),
you have to edit script /usr/local/GSUNY/GPARAM (or \$USER/.gsunyrc) as well.

\section{Results and discussion}

An execution time for different scripts is shown in table \ref{submission}. Here we have to emphasize
that this time is required only to submit the job to the standard job queue. We use LSF at BNL and
Open PBS at SUNYSB as local job managers.

\begin{table}[t]
\begin{center}
\caption{Measurement results}
\begin{tabular}{|l|r|c|}
\hline \textbf{Command} & \textbf{Execution time} & \textbf{Remarks} \\
\hline gping     &  6 secs &   \\
\hline gsub      & 42 secs &   \\
\hline gsub-p    & 26 secs &   \\
\hline gsub-s    &  5 secs &   \\
\hline qsub-unm  & 11 secs &   \\
\hline gsub-data & 17 secs &   including \\
 & & looking \\
 & & at the catalog \\
\hline
\end{tabular}
\label{submission}
\end{center}
\end{table}

By looking at the table it is easy to realize that job submission takes time. It is not surprising
because we used Globus command {\it globus-job-run} to get current load of the cluster and we did not
use Globus Index Information Service (GIIS) \cite{mds}. Also we use the parameter {\it -stage} for the
command {\it globus-job-submit} to copy a job script from local desktop to remote cluster (it takes
time as well).  Somebody may think that there is no reason to submit the job with expected execution
time 1 minute because the overheads for job submission might be more than execution time.  On the 
other hand
if you do not know exactly the situation on clusters your short 1 minute job may stay in input queue
many hours due to high load or some problem on the computing cluster.

To decrease the delays we plan to use Monitoring and Discovery Service (MDS)  
\cite{mds} in nearest time.

\section{CONCLUSION}

The first working prototype includes central PHENIX (BNL), SUNYSB, Vanderbilt University (only
replication subsystem for PHENIX file catalog was deployed), University of New Mexico \cite{unm} (only simple
job submission scripts were deployed).  All components are working and
results look promising:

\begin{itemize}

\item The distributed information about location of significant collaboration physics data is
delivered in uniform and consistent manner;

\item Job submission from remote universities could be directed to less loaded cluster where required data
are. It helps to use computing resources more effectively in two different scenarios.

\begin{itemize}

    \item In first scenario it is needed to collect all available distributed computing power to do an 
analysis. In this
          situation you need to distribute the data around computing clusters before starting the job 
          chain. Described tools will help to load all available clusters evenly.

    \item In second scenario there are already distributed over several clusters data (experimental and 
simulated). In this case the using of described tools will help
          to minimize data moving over WAN.

\end{itemize}

\end{itemize}

Apparently it was done the first step in implementing the flexible distributed computing 
infrastructure. Some additional work is required on robust interfaces in between cataloging 
engine, job submission tools, job accounting tools, data moving, and trace information about the 
jobs.

Finally, during the implementation we learned several lessons:

\begin{itemize}

\item Deployment of the Grid infrastructure in collaboration scale may not be a business for one
person.

\item Grid architecture has to be supported at central computing facilities.

\item Better understanding of our needs in Grid comes with real deployment of the components.

\end{itemize}

\begin{acknowledgments}

Authors have to mention the people who gave us valuable information and 
spent discussion time: Rich Baker, Predrag Buncic, Gabriele Carcassi, Wensheng Deng, 
Jerome Lauret, Pablo Saiz, Timothy L. Thomas, Torre Wenaus, Dantong Yu. Special thanks for our colleagues Nuggehalli N. 
Ajitanand, Michael Issah, Wolf Gerrit Holzmann who asked many questions and helped to formulate 
things more clearly.

\end{acknowledgments}

The work was done with support of the grant NSF-01-149 (award number 0219210).



\end{document}